# Study on the power scaling and mode instability property in an all-fiber narrow-linewidth tapered fiber amplifier

**Long Huang[1,2], Zichao Zhou[1,2], Chen Shi[1,2], Rumao Tao[1,2], Xiaolin Wang[1,2], Pu Zhou[1,2]**

[1] College of Optoelectronic Science and Engineering, National University of Defense Technology, Changsha, Hunan, 410073, People's Republic of China.

[2] Hunan provincial collaborative innovation center of High Power Fiber Laser, Changsha, Hunan, 410073, People's Republic of China.

Email: zhoupu203@163.com

**Abstract**

We develop a high-power narrow-linewidth fiber amplifier based on a tapered Yb-doped fiber. In the experiment, the narrow end of the tapered fiber is used as the input port of both pump source and signal light to form a robust all-fiber configuration. The stimulated Brillouin scattering (SBS) is effectively suppressed owing to its gradually increased mode area as well as continuously changed Brillouin frequency shift dependent on the core diameter. As a result, an output power of 260 W without SBS and mode instability (MI) is obtained, with a slope efficiency of 71.3 % and a narrow linewidth of ~2GHz. However, the MI rather than SBS is observed when output power reaches 260 W. The features of MI are experimentally studied in detail. It is pointed out that, the MI seems the primary and durative limitation factor of the tapered fiber amplifier for high-power narrow-linewidth output, although it has advantage to reduce other nonlinear effect. To the best of our knowledge, this is the first detailed experimental study on MI in a tapered fiber. At last, some worthwhile discussion about the optimization of the tapered fiber and the whole system for both SBS and MI suppression is conducted based on the present and previous results.

## 1. Introduction

It is well known that the Yb-doped fiber lasers (YDFLs) have wider and wider applications in various fields due to their pronounced advantages such as high efficiency, outstanding beam quality, free thermal management, robust structure, and so forth [1]. Benefitting from the rapid improvement of the high-power laser diode (LD) and large-mode-area (LMA) double cladding active fiber, the output power of the near-diffraction-limited YDFLs have experienced a significant leap to tens of kilo-watts in the past decade [2-4]. The high-power narrow-linewidth YDFLs have also reached the level of several kilo-watts [5-7]. However, the output power of the narrow-linewidth YDFLs is generally lower than that of the broadband ones due to the limitation of nonlinear effects, especially the stimulated Brillouin scattering (SBS) effect which is considered as the primary limitation factor [8]. One practical method for SBS suppression is to enlarge the mode area of the active fiber [9]and shorten its length by adopting high doping concentration[10]. Nonetheless, with increased fiber size, the active fiber is able to support transverse modes, which results in another serious problem—thermal-induced mode instability (MI) [11-12]. For the time being, the MI gradually becomes a new research focus and is regarded as a tough obstacle that we have to overcome for further power scaling of narrow-linewidth fiber laser [13-14].

Remarkably, due to core-diameter-dependent Brillouin frequency shift, the so-called tapered Yb-doped fiber (T-YDF) with gradually changed core diameter in longitudinal direction has been demonstrated with capacity of SBS suppression, which is beneficial for power scaling of the narrow-linewidth YDFLs [15-16]. If a large core size is also adopted in the design process of the T-YDF, the suppression effect on SBS can be further enhanced due to gradually increased mode area in the longitudinal direction. Except SBS suppression, the T-YDF has been demonstrated to have many other attractive advantages such as high cladding pump absorption coefficient due to mode mixing effect[17], high beam quality due to mode selection effect [18-20], high pump conversion efficiency[21], spontaneous amplified emission (ASE) mitigating property[22-23], brightness enhancement characteristic [24], self-pulse suppression [21, 23], generation of special wavelength [25] and so on. By adopting a T-YDF, a record broadband output of 750 W and single-frequency output of 160 W have been obtained both with high beam quality. It is anticipated that further power scaling can be realized by enlarging the core diameter of the T-YDF to improve the energy storage capacity [19]. However, most of the reported work have adopted free-space and counter-propagating pumping onto the wide end of the T-YDF. By contrast, an all-fiber configuration may be more convenient and reliable. Besides, it is noted that, as a limitation factor for high-power fiber laser, the MI in the T-YDF laser has not been studied in details up to now. Some valuable experimental results about MI have been mentioned in [26], but the T-YDF laser in this work belonged to an oscillator, which could not reflect the property of a T-YDF-based amplifier. Moreover, the details of experimental setup and corresponding results are not provided.

In this paper, a high-power narrow-linewidth T-YDF-based amplifier based on all-fiber configuraion is developed. As a result, an output power of 260 W without SBS and MI is obtained, with a slope efficiency of 71.3 % and a narrow linewidth of ~2GHz. When the output power is above 260 W, the MI rather than SBS is observed. The overall features of MI are experimentally studied. It is pointed out that, by using the narrow end as the injection port of both pump source and signal laser to achieve an all-fiber configuration, the MI may be the primary and durative limitation factor of a T-YDF-based amplifier. To the best of our knowledge, it is the first detailed experimental study of MI in a T-YDF amplifier. Finally, some future optimization of the T-YDF and the whole system for both SBS and MI suppression is discussed based on the present and previous results.

## 2. Experimental Setup

The experimental setup is depicted in Fig. 1(a). The laser system has a master oscillator power amplifier (MOPA) configuration, consisting of a seed laser and three amplifiers in all-fiber structure. The center wavelength of the single-frequency seed laser is ~1064 nm, possessing a linewidth of 20 kHz and an adjustable output power within 100 mW. In the experiment, the output of the seed laser is selected as 55 mW. The single-frequency seed is phase-modulated to be a narrow-linewidth one by an arbitrary function generator and phase modulator. The amplitude and frequency of sinusoidal modulation signal is adaptable according to the requirement. In the first-stage amplifier, the 2.5 m-long Yb-doped single-cladding Yb-doped fiber has a core diameter of 6 μm and an inner cladding diameter of 125 μm, bidirectionally pumped by two 500 mW-level laser diodes (LDs) at 974 nm via two wavelength division multiplexes. After the first-stage amplifier, the boosted signal with a power of 200 mW is injected into the second-stage amplifier after a bandpass filter and an isolator. In the second-stage amplifier, the double cladding active fiber has a length of 3.5 m, core/inner cladding diameters of 10/125 μm and a cladding-pump absorption coefficient of ~4.8 dB/m at 976 nm. Pumped by two 55 W-level LDs at 976 nm, the signal is amplified to be ~ 6.6 W and launched into the main amplifier after an isolator and a tapper. By the tapper, about 0.01 % signal power of the second-stage amplifier is guided into a Fabry-Pérot interferometer (FPI) to monitor the modulation quality of the seed laser. Meanwhile, about 0.01% backward light of the main amplifier is monitored by a power meter to judge the occurrence of SBS.

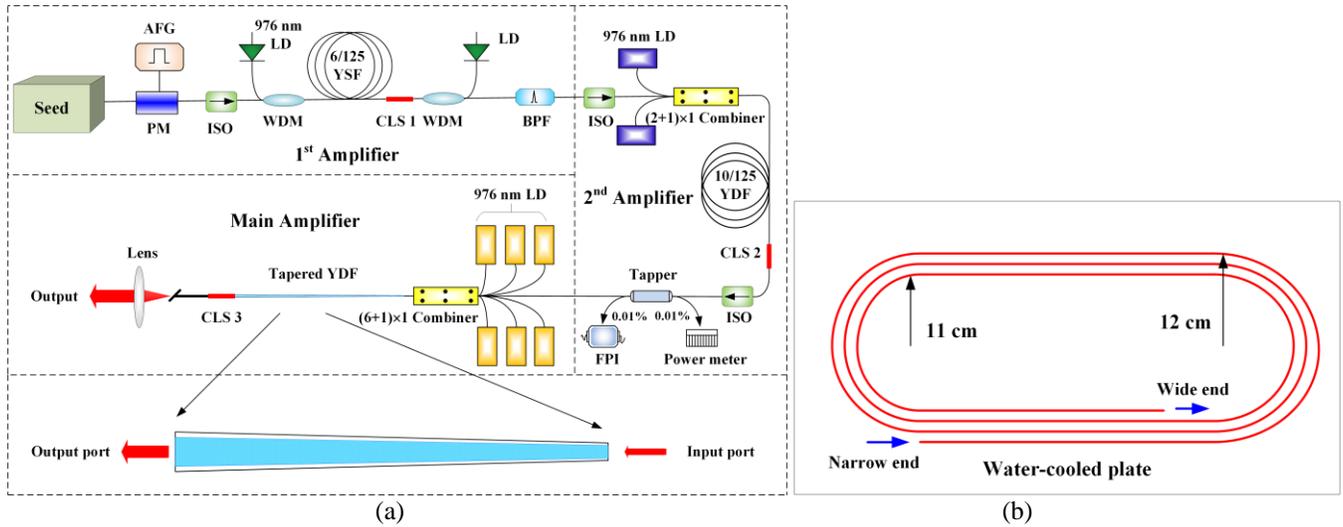

**Fig. 1** Experimental setup and (b) arrangement sketch of the T-YDF on a water-cooled mental plate. (PM: phase modulator, AFG: arbitrary function generator, ISO: isolator, LD: laser diode, WDM: wavelength division multiplex, YSF: Yb-doped single-cladding fiber, CLS: cladding light stripper, BPF: bandpass filter, YDF: Yb-doped double-cladding fiber, FPI: Fabry-Pérot interferometer, DM: dichromic mirror)

In the main amplifier, a piece of 7.2 m-long homemade T-YDF is used as the gain medium and pumped by six 95 W-level LDs at 976 nm. The signal port of the (6+1)×1 combiner has core/cladding diameters of 15/130 μm and the output port has core/cladding diameters of 15/250 μm. The T-YDF has core/inner diameters of 20.4/237.1 μm at the narrow end and 46.9/579.9 μm at the wide end respectively, corresponding to a core taper ratio of 46.9/20.4=2.3. The core numerical aperture (NA) is ~0.06 and the cladding pump absorption coefficient is ~1.4 dB at 976 nm. To realize a robust all-fiber configuration, the signal and pump power are both injected into T-TDF via its narrow end in particular, which helps to avoid the pump leakage caused by vignetting effect [19] compared with injecting pump source from the wide end. What's more, this configuration also benefits suppressing SBS owing to its gradually increased mode area in the propagation direction of laser. To arrange and cool the T-YDF conveniently, it is placed in the V-grove of a water-cooled mental plate as depicted in Fig. 1(b), with a minimum coiling radius of ~11 cm in the internal ring and a maximum coiling radius of ~12 cm in the external ring. After the main amplifier, a 0.5 m-long delivery fiber with core/caldding diameters of 48/400 mm is spliced to launch the output power and angle-cleaved to avoid feedback as well as end damage. After each amplifier, the residual cladding light is removed by the cladding light stripper comprising of index gel, including the main amplifier.

## 3. Results and Discussion

3.1 Power scaling of the all-fiber narrow-linewidth T-YDF-based amplifier

In the experiment, a narrow-linewidth signal laser is obtained by imposing a sinusoidal modulation signal on the single-frequency seed laser. The frequency and amplitude of the sinusoidal modulation signal are 80 MHz and 10 V respectively. Fig. 2 shows the oscilloscope scanning image of the seed laser captured by FPI before and after modulation. The FPI has a free spectral range (FSR) of 4GHz and a fineness of 10 MHz. It is noted that, after modulation, the seed laser has a narrow linewidth of ~2GHz (~0.0066 nm). According to the reported work in Ref. [27]and [28], the SBS threshold in this T-YDF-based amplifier is estimated to be above 350 W.

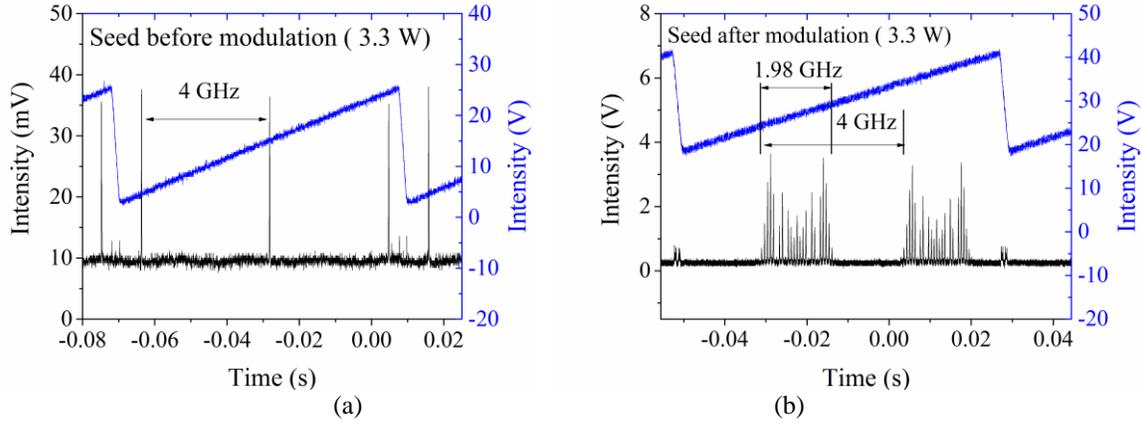

**Fig. 2** The oscilloscope scanning image of the seed laser captured by FPI (a) before and (b) after modulation

After the amplification in two pre-amplifiers, the modulated seed laser is boosted to be ~6.6 W and then launched into the T-YDF-based main amplifier. For the main amplifier, the dependence of the output power on the pump source is shown in Fig. 3(a). Below 260.6 W, the output power increases linearly with a slope efficiency of 71.3 %. However, when the output power is beyond this level, its increase trend slow down and the slope efficiency decreases to be 52.5 % until the pump-limited maximum output power of 318 W is reached. This is caused by the occurrence of MI which will be discussed below. As shown in Fig. 3(b), the backward light of the main amplifier increases linearly throughout the power scaling process, indicating that there is no SBS. Thus, we can conclude that not SBS but MI is the limitation factor for further power scaling in present condition.

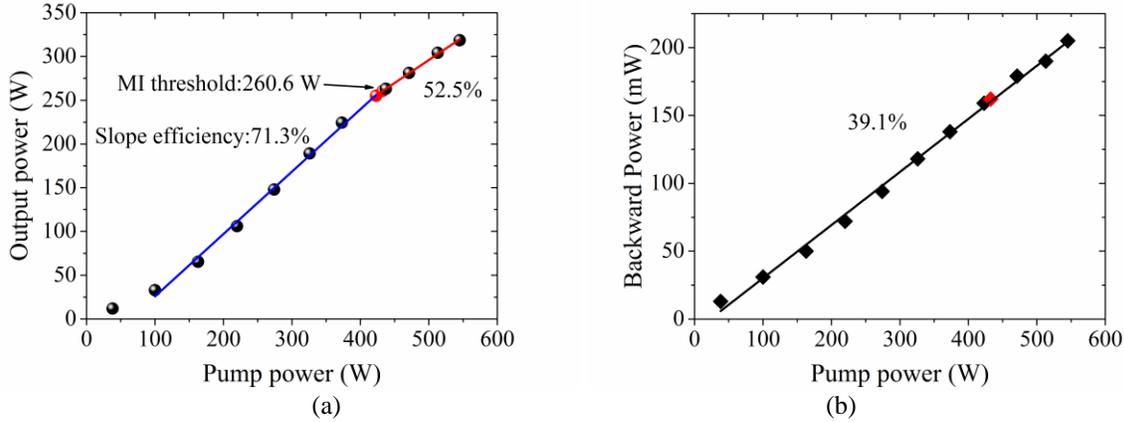

**Fig. 3** The (a) output power and (b) backward power of the main amplifier vs pump power.

The Fig. 4(a) shows the spectrum at the maximum output power. It is noted that there is a little residual pump light and ASE in the output. It is thought that the ASE is induced by the ASE noise originated from the pre-amplifier due to saturated gain. However, the intensity of the signal light is more than 33 dB higher than that of the residual pump light and ASE, so they can be neglected. The spectral details monitored by an optical spectrum analyzer (OSA) are shown in Fig. 4(b) with a measurement resolution of 0.02 nm. The measured 3 dB bandwidths are always around 0.03 nm, indicating that there is no spectral broadening in the main amplifier.

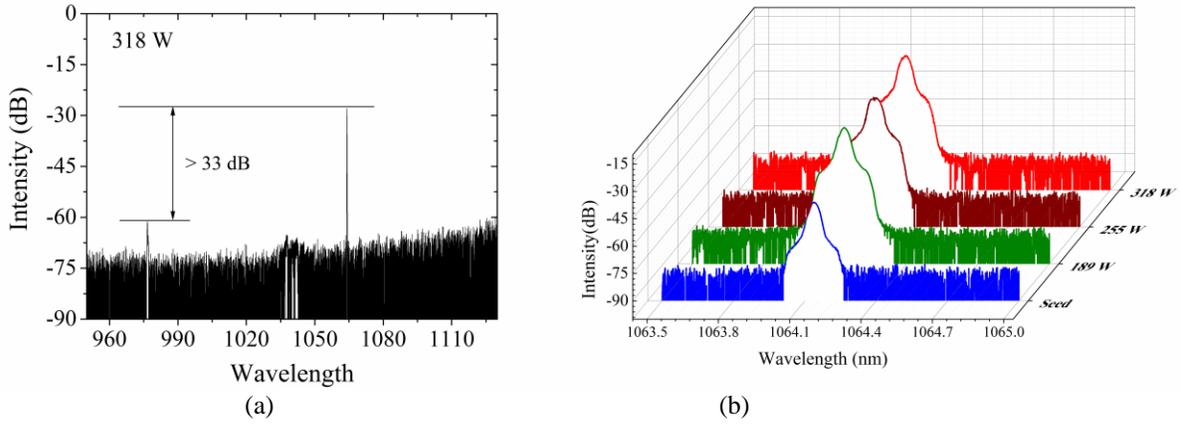
**Fig. 4** (a) the spectrum at output power of 318 W and (b) the spectral evolution of the main amplifier in detail.

Due to the resolution limitation of OSA, we directly couple a little part of output light into the FPI to scan the accurate linewidth of the laser. As shown in Fig. 5, at output power of 255 W and 318 W, the measured linewidths are 2.01 GHz and 1.98 GHz respectively. Compared with the linewidth of the seed laser (2 GHz), spectral broadening is hardly observed. These results effectively demonstrate the capacity and potential of the T-YDF for high–power narrow-linewidth amplification again.

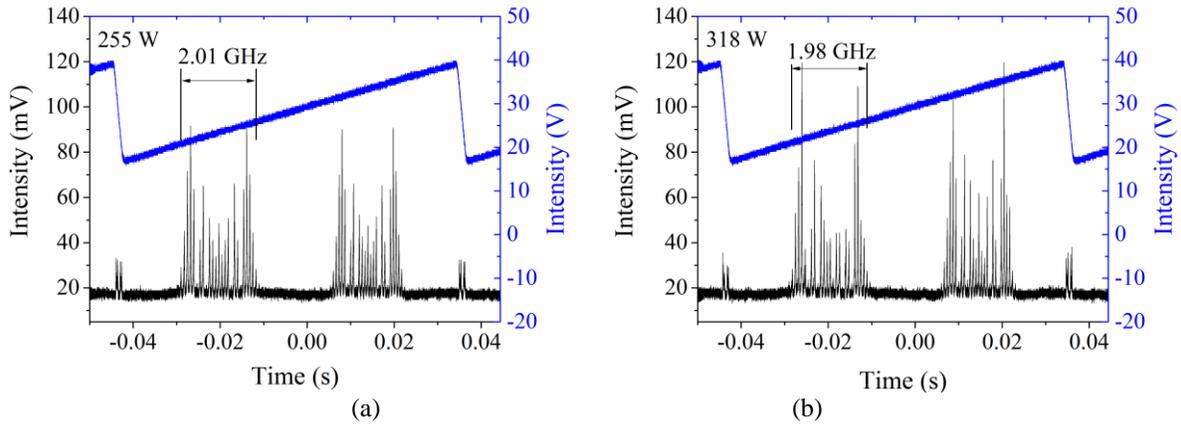
**Fig. 5** The oscilloscope scanning image of the laser captured by FPI at output power of (a) 255W and (b) 318 W respectively.

3.2 Observation of MI in the all-fiber narrow-linewidthT-YDF-based amplifier

As mentioned above, the MI is observed as the obstacle for power scaling. To reveal the property of MI in the T-YDF, we experimentally study the MI in details. As the output power increases, the time traces of the laser and corresponding Fourier transforms are captured by a photo-detector and an oscilloscope [29]. As shown in Fig. 6, when the output power increases to 260.6 W, the time trace image of the laser begins to exhibit sawtooth-like peaks, just corresponding to inflection point (260.6 W) of power scaling mentioned above. Moreover, the sawtooth-like peaks become more and more obvious with output power increasing. This is a typical feature of the occurrence of MI and thus we regard the power level of 260.6 W as the MI threshold of the laser.

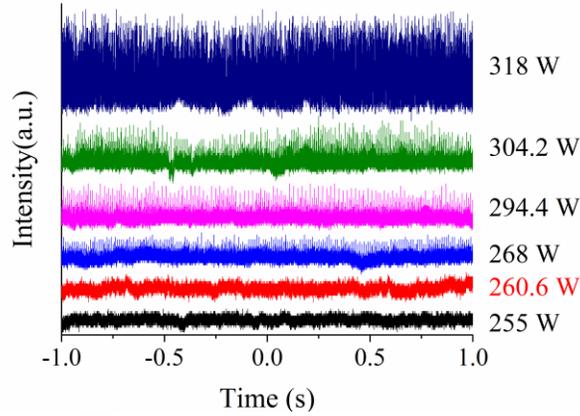
**Fig. 6** The time trace images of the laser at different output power.

Accompanying the sawtooth-like peaks in time domain, the frequency spectra of the time traces by Fourier transform also display some apparent features. Before the onset of MI, the frequency spectrum distributes uniformly without any dominant frequency component as shown in Fig. 7(a). After the onset of MI, the frequency components within the range of 0~7 kHz show up, mainly locating at region around 2.5 kHz, as shown in Fig. 7(b). These frequency components in kHz level also represent a typical feature of MI in fiber laser as reported in [13-14, 30], verifying the fluctuation in time domain. However, it is noted that, there is no fixed or single-dominant characteristic frequency components as reported in Ref.[14] at any power levels. Namely, the dominant frequency components are always distributed in a region. Besides, with output power increasing, the frequency components in low-frequency region become stronger and stronger than those in high-frequency region, as shown in Fig. 7(b)-(f). These features indicates that the MI-induced modes coupling process in this T-YDF may be much more complicated than that happening in a traditional uniform LMA fiber in which only a few modes are supported along the whole fiber.

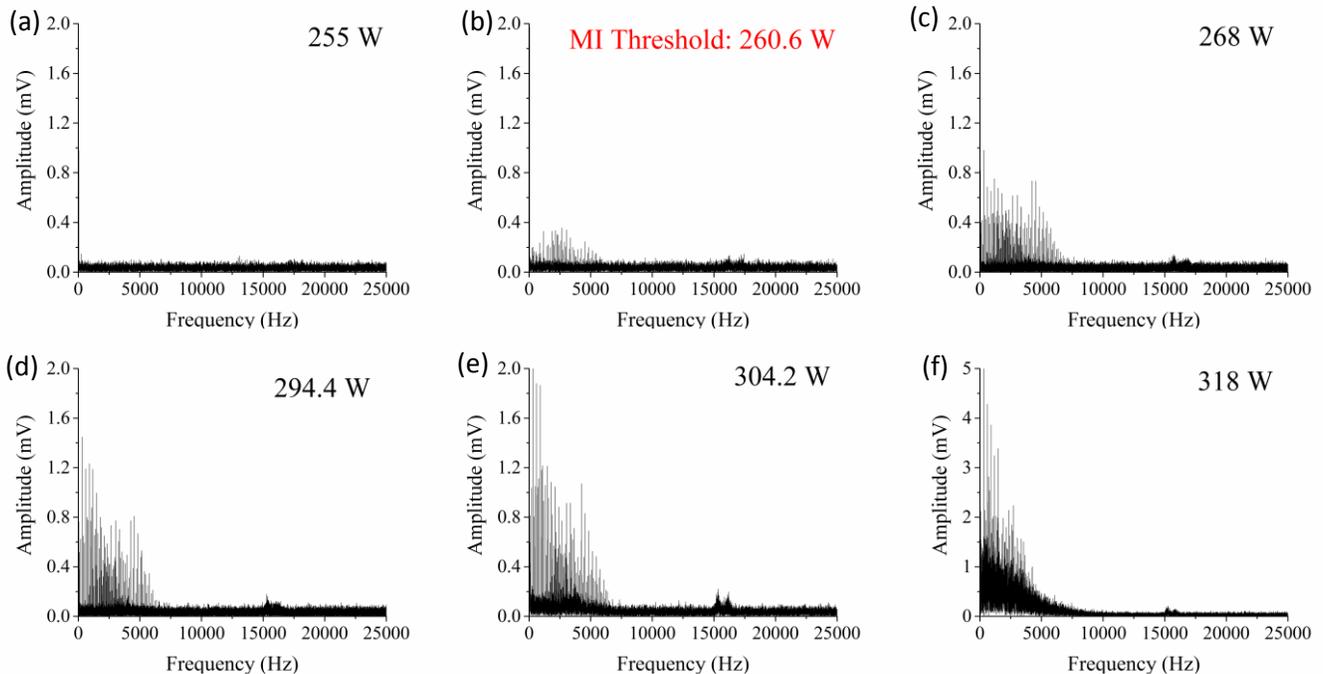
**Fig. 7** The Fourier spectra of the time trances (a) before and (b)-(f) after the onset of MI.

To evaluate the MI threshold quantitatively, we also define a character factor according to the Fourier spectrum of the time trace [13], shown as:

$$\sigma = \left[\frac{\int_0^{12.5k} P(v)dv}{\int_{12.5k}^{25k} P(v)dv} - 1\right] \times 100\% \tag{1}$$

where $P(v)$ is the power intensity of the component with frequency of $v$. Below the MI threshold, the $P(v)$ in the range of 0-12.5 kHz are nearly equal to those in the range of 12.5-25 kHz. Thus, the value of $\sigma$ is around zero, indicating that the MI does not occur. When the MI arouses, the $P(v)$ in the range of 0-12.5 kHz are much bigger than those in the range of 12.5-25 kHz. So, the value of $\sigma$ increases rapidly, indicating the onset of MI. Based on this fact, we quantitatively define the MI threshold as the power level at which the value of $\sigma$ is around 10 %. The Fig. 8 shows the values of $\sigma$ at different output power. It is noted that, at the output power of 260.6 W, the value of $\sigma$ close to 10 %, so it is reasonable to regard 260.6 W as MI threshold. This result exemplifies that the definition of MI threshold based on the Fourier spectral distributions of time traces is in good agreement with the occurrence of sawtooth-like fluctuation in time domain and the inflection point of power increase. By calculation, with a further power scaling of 58 W above MI threshold, the value of $\sigma$ explosively increases to be even 6752 % at the output power of 318W, which is much more remarkable than that observed in a 30/250 μm uniform LMA fiber in Ref.[13], in which the value of $\sigma$ only increases to be ~90 % under similar power scaling range above MI threshold. This phenomenon also demonstrates that the mode coupling process after the onset of MI is stronger than a uniform LMA fiber with a moderate core diameter.

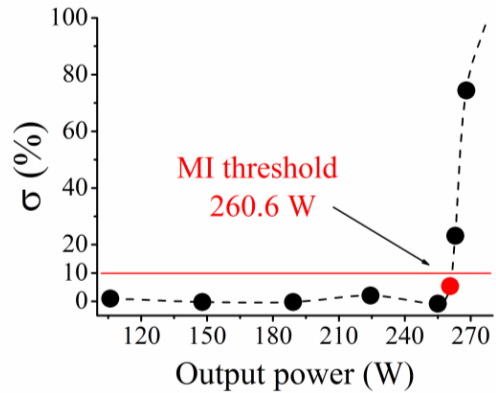

**Fig. 8** The value of the characteristic factor $\sigma$ at different output power.

To illustrate the relationship between mode content and MI, the beam profile at different output power are measured and some results at typical power levels are shown in Fig. 9. It is noted that, there seems a little cladding signal light in the seed laser which is not absolutely stripped by the cladding light stripper (CLS). The possible reason is that, when the signal light propagates from the narrow end to the wide end, the effective numerical aperture (NA) of both fiber core and inner cladding are improved due to gradually increased diameter profile[17], so it becomes more difficult to strip the signal light leaking into the inner cladding. As known to us, compared with the fundamental mode, it is easier for high-order modes to be coupled into the cladding, so the existence of cladding signal light indicates that there may be much high-order modes in the core. Actually, due to large core diameters of 20.4 μm and 46.9 μm at the narrow and wide end of this T-YDF, there are theoretically several and even tens of transverse modes supported respectively. Besides, the T-YDF is coiled with a radius of 11 cm in the internal ring and a radius of 12 cm in the external ring respectively. These coiling radiuses are far smaller than that discussed in Ref.[21], so some high-order modes may be excited by the coiling state. Moreover, due to all-fiber configuration, some high-order modes may be excited by the splicing point at the narrow end and then propagate along the fiber until launch into free space. It is thought that the excited high-order modes may be the reason why the T-YDF amplifier has a relatively low MI threshold of 260.6 W and more complicated mode coupling process than a uniform LMA fiber, as discussed above.

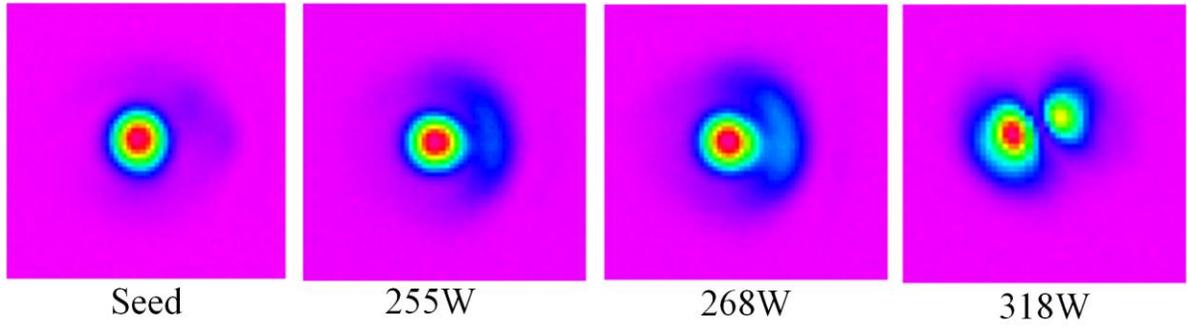

**Fig. 9** The beam profiles and M² results at some typical power levels.

As shown in Fig. 9, the cladding signal light increases gradually with the output power increasing, indicating the increase of high-order modes content in the core. Different from the few-modes LMA fiber as discussed in [11], no split of beam profile is observed when the output power is at the vicinity of MI threshold (268 W), which is due to the FPS (frames per second) of the camera is far slower than the coupling frequency of MI. At the output power of 318 W, obvious split of beam profile is observed, which is a typical feature of MI and indicates that considerable power is coupled into high-order modes by MI. Compared with a uniform LMA fiber, these results sufficiently demonstrate complex mode content and mode coupling process in this T-YDF, as illustrated by the Fourier spectra and the value of σ.

Besides, it is noted that the slope efficiency of this T-YDF-based amplifier is relatively low compared with the typical results of uniform LMA-based amplifiers [5, 10]. This result indicates that maybe more pump energy is converted to the thermal load for reasons such as relatively high background loss of this T-YDF, which may be another reason why the MI threshold of this T-YDF-based amplifier is relatively low.

3.3 Effect of the signal linewidth on MI

In this section, in order to verify whether the MI threshold is linked to the linewidth or not, the modulation signal imposed on the single-frequency seed is changed to be 100 MHz and 36 V by adopting the driver source of an acoustic-optical modulator. With increased modulation voltage and frequency, the scanning linewidth of the seed laser under new modulation is shown in Fig. 10(a), exceeding the FSR (~4GHz) of FPI. However, we can estimate the linewidth to be ~4.9 GHz (~0.016 nm, out of the resolution of OSA) based on the reported work in Ref.[31]. It is expected that the SBS threshold can be further increased by broadening the linewidth of the seed laser, which is in favor of obtaining higher narrow-linewidth output power.

It is noted from Fig. 10(b) that, neglecting the resolution limitation of OSA, there is still no spectral broadening observed in the main amplifier, which exemplifies the sustainable capacity of the T-YDF for high-power narrow-linewidth amplification. However, as shown in Fig. 10(c), the slope efficiency decreases from 71.3 % to 53.6 % with an inflection point at 261.2 W and a pump-limited output power of 318.5 W, which is almost identical with the results before broadening the linewidth of seed laser. Neglecting the measurement error, in can be thought that the MI threshold does not vary before and after changing the modulation parameter.

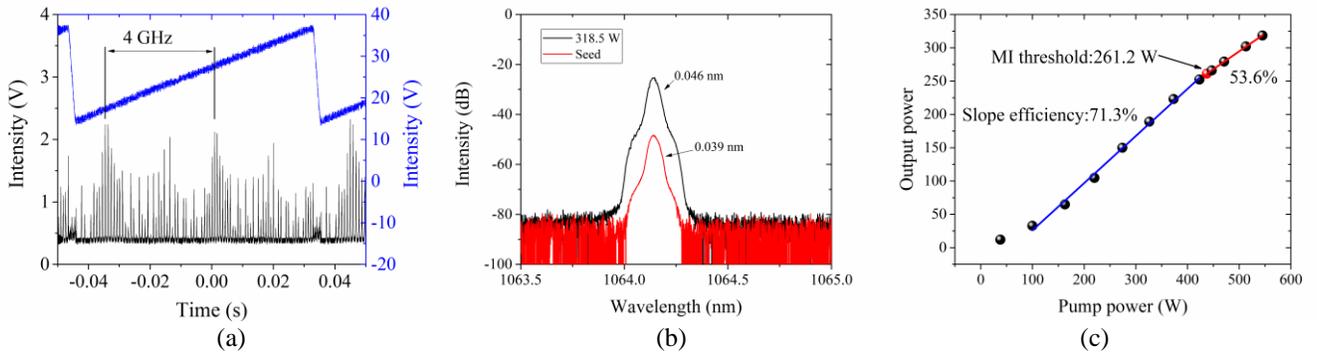

**Fig. 10** The (a) scanning linewidth of the seed laser, (b) spectral evolution and (c) output power of the main amplifier vs pump power after broadening the linewidth of seed laser.

Besides, the Fourier spectra of the time traces are also analyzed, as shown in Fig.11(a)-(c). The frequency components that show up after the onset of MI are also distributed in the range of 7.5 kHz. The definition of MI threshold according to the value of σ is also in agreement with the inflection point of power increase. Moreover, there is also no fixed or single-dominant characteristic frequency component. The low-frequency components also become stronger with the output power increasing.

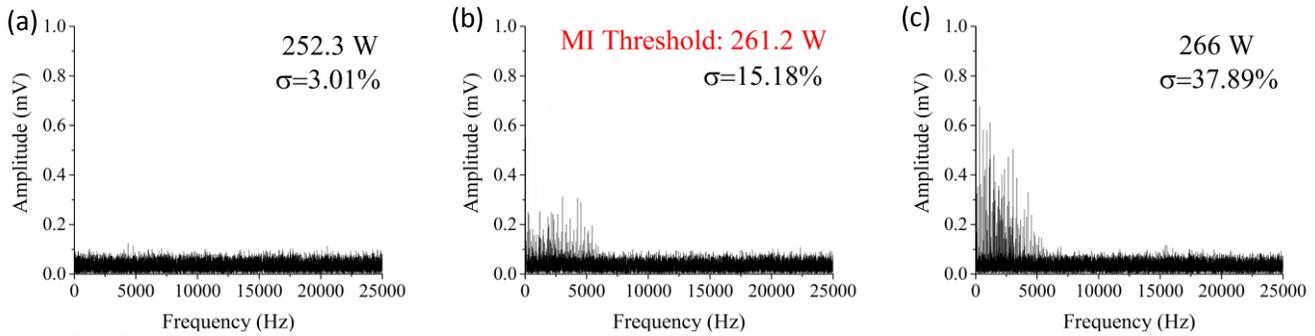

**Fig. 11** The Fourier spectra of the time traces of the main amplifier under new modulation parameter.

It is reported in [32-33] that, the MI threshold can be increased under broad lindwidth of the signal light, on condition that the coherence time of signal light is less than the group delay of two transverse modes. In our experiment, the reason why the MI threshold is not improved under broadened linewidth of the signal light may be that the linewidth is not wide enough. However, if the linewidth of the signal light is intentionally broadened enough to achieve an increase of MI threshold such as on the order of several nanometers [33], the linewidh of output laser may exceed the narrow-linewidth category. To sum up, it seems that not SBS but MI is the primary and durative limitation factor of a T-YDF for high-power narrow-linewidth output, although it has a tunable capacity and potential for SBS suppression by reasonably changing the linewidth of seed laser. So, one must manage to improve the MI threshold when constructing a T-YDF-based laser for high-power narrow-linewidth output.

3.4 Overall discussion about the T-YDF for high-power narrow-linewidth output based on a robust all-fiber configuration

The T-YDF has been reported to possess intrinsic SBS suppression property [15-16] and ASE mitigation property [22-23]. However, the T-YDF also has some special advantages and corresponding disadvantages determined by how we make the best of a T-YDF. Namely, if we want to construct a high-power narrow-linewidth fiber laser with high beam quality, high MI threshold and a robust all-fiber configuration, we need to consider the propagation direction of both pump source and signal laser, so as to make the best of its advantages and avoid its disadvantages.

It is known that, if the pump source is launched into the T-YDF by its wide end, the pump absorption can be enhanced by mode mixing effect due to gradually-decreased diameter of the inner cladding in the pump propagation direction [19], which is beneficial for reducing the fiber length and making use of the low-brightness pump source. However, due to too large diameter of the wide end, this pump injection method generally depends on a bulky configuration, so sometimes it may be not suitable for an all-fiber configuration. More importantly, this pump injection method needs to avoid pump leakage induced by vignetting effect and thus the numerical aperture of pump source must be limited in a sufficiently low level, which also relies on a free-space reshaping system for the low-brightness pump source. So, sometimes it is necessary to consider using the narrow end of a T-YDF to build an all-fiber configuration. In this situation, the cladding absorption coefficient of pump source will reduce a little because of the absence of mode mixing effect, but we can avoid the pump leakage induced by the vignetting effect. The efficient pump absorption can be ensured by properly adding the fiber length. If a T-YDF with relatively high doping concentration is adopted, we need not worry about the fiber length at all. As proved in our work, an excellent slope efficiency of 71.3 % can also be achieved by launching the pump source into the narrow end of a T-YDF based on an all-fiber structure.

Except the injection direction of pump source, the output direction of signal laser is also significant. It is known that, the narrow end of a T-YDF is functional to select the core modes, which is beneficial for improving the beam quality and brightness of the output power [21, 24]. So, it is wise to use the narrow end as output port for the pursuit of high beam quality. However, compared with using the narrow end as the output port, it also has particular advantages to use the wide end as the output port, such as higher self-pulsing threshold, higher beam contrast (the ratio of beam propagating in the fiber core) and higher pump conversion efficiency [21, 24]. Moreover, it is expected that using the wide end as output port also helps to reduce other nonlinear effect (including SBS) due to gradually-increased mode area in the propagation direction of signal laser. So, the most important issue about using the wide end as output port is the beam quality and MI threshold. Fortunately, no matter which end of a T-YDF is used as output port, the mode selection function of the narrow end does work all the same[21]. In other words, if only fundamental mode or a few modes are supported in the narrow end, the wide end can also output a laser with near-diffraction-limited beam quality, as proved in [17, 20]. But then, one must also pay careful attention to possible excitation of high-order modes induced by coiling or local bend of the T-YDF, especially when the local bend is located near the narrow end while the wide end is selected as output port, in which more power will be coupled into high-order modes because the small-diameter section is more sensitive to coiling or local bend and the excited modes may further excite higher-order modes as they propagate from the small-diameter section to the wide end [21]. The excited high-order modes not only degrade the beam quality, but also may promote the modes coupling after the onset of MI and even decrease the MI threshold, as discussed in our experiment. Likewise, a relatively degraded beam quality with $M^2$=2.7 is obtained in [22], mainly due to too large core diameter of 15 μm in the narrow end and possible coiling of the T-YDF in an all-fiber configuration.

Based on the matter discussed above, it can be concluded that, with the T-YDF arranged in a reasonable coiling or bend state, a crucial factor which determines the beam quality and may be linked to MI threshold of output power from the wide end is the number of modes supported in the narrow end, i.e. the diameter of the narrow end. For an oscillator based on the T-YDF, the mode selection function can be enhanced due to two-path transmission through the small-diameter section. For an amplifier, the mode selection function may be a little weaker because the laser transmits the small-diameter section for only one time, which may make a difference in terms of beam quality and MI threshold, especially when the wide end is used as the output port. So, for an amplifier, selecting a relatively small diameter for the narrow end and avoiding the excitation of high-order modes seems more important. In fact, if we want to construct a robust all-fiber T-YDF-based MOPA, it is better to select the wide end as the output port because of its generally too large diameter, so as to ensure the compatibility between T-YDF and other all-fiberized components such as pump combiner. Correspondingly, the narrow end is naturally used as the injection port of both pump source and signal laser. So, the splicing quality at the narrow end and reasonable coiling or bend state are very important for mitigating excited high-order modes, which ensures high beam quality and may help to increase the MI threshold.

In summary, the T-YDF has intrinsic SBS-suppression property for high-power single-frequency or narrow-linewidth output. By choosing the narrow end of a T-YDF as the injection port of both pump source and signal laser, it is proved feasible and efficient to build a robust all-fiberized high-power MOPA with single-frequency or narrow-linewidth property, high beam quality, on condition that a relatively small diameter (a single-mode one is best) is selected for the narrow end and the excitation of high-order modes are successfully avoided by improving the splicing quality and optimizing the coiling or local bend state of the T-YDF, which may also benefit improving MI threshold. When the diameter of the wide end is fixed, a small core diameter for the narrow end means a high tapered ratio (the ratio of core diameters at wide and narrow end) in the manufacture process. If a low core NA is employed, the beam quality and MI threshold can be improved further [34]. Furthermore, the reduction of thermal load by methods such as improve the manufacture quality of the T-YDF to reduce the background loss may also help to improve the MI threshold. It is better if the longitudinal diameter profile can also be optimized as discussed in [24].

## 4. Conclusion

We demonstrate a high-power narrow-linewidth MOPA based on a T-YDF. The narrow end of the T-YDF is used as the injection port of both pump source and signal laser to conveniently build a robust all-fiber configuration. The SBS is effectively suppressed owing to gradually increased mode area in the T-YDF as well as continuously changed Brillouin frequency shift dependent on the core diameter. As a result, an output power of 260 W without SBS and MI is obtained, with a slope efficiency of 71.3 % and a narrow linewidth of ~2GHz. However, the MI rather than SBS is observed when the output power is above 260 W. The overall features of MI are experimentally studied. It is pointed out that the MI may be the primary and durative limitation factor of the T-YDF-based amplifier for high-power narrow-linewidth output, although it has advantage to reduce other nonlinear effect. To the best of our knowledge, it is the first detailed experimental study on MI in a T-YDF. Further, some discussion about optimization of the T-YDF and the whole system for both SBS and MI suppression is conducted, which is referable when one attempts to further explore a T-YDF laser for high-power narrow-linewidth output based on all-fiber configuration.

## Acknowledgements

The authors would like to acknowledge Professor Zhiyong Pan for the supply of the tapered YB-doped fiber. This work was supported by the National Natural Science Foundation of China under Grant No. 61322505 and the foundation for the author of National Excellent Doctoral Dissertation of China (FANEDD, No. 201329).

**References:**

# References:

[1]     Jauregui C, Limpert J and Tunnermann A 2013 High-power fiber lasers *NAT PHOTONICS* **7** 861-7
[2]     Yu H, Zhang H, Lv H, Wang X, Leng J, Xiao H, Guo S, Zhou P, Xu X and Chen J 2015 3.15 kW direct diode-pumped near diffraction-limited all-fiber-integrated fiber laser *Appl. Opt.* **54** 4556-60
[3]     Stiles E 2009 New developments in IPG fiber laser technology. In: *Proceedings of the 5th International Workshop on Fiber Lasers,* pp 4-9
[4]     Fang Q, Shi W, Qin Y, Meng X and Zhang Q 2014 2.5 kW monolithic continuous wave (CW) near diffraction-limited fiber laser at 1080 nm *LASER PHYS LETT* **11** 105102
[5]     Beier F, Hupel C, Nold J, Kuhn S, Hein S, Ihring J, Sattler B, Haarlammert N, Schreiber T, Eberhardt R and Nnermann A T U 2016 Narrow linewidth, single mode 3 kW average power from a directly diode pumped ytterbium-doped low NA fiber amplifier *OPT EXPRESS* **24** 6011-20
[6]     Ma P, Tao R, Su R, Wang X, Zhou P and Liu Z 2016 1.89 kW all-fiberized and polarization-maintained amplifiers with narrow linewidth and near-diffraction-limited beam quality *OPT EXPRESS* **24** 4187-95
[7]     Xu Y, Fang Q, Qin Y, Meng X and Shi W 2015 2 kW narrow spectral width monolithic continuous wave in a near-diffraction-limited fiber laser *Appl Opt* **54** 9419-21
[8]     Dawson J W, Messerly M J, Beach R J, Shverdin M Y, Stappaerts E A, Sridharan A K, Pax P H, Heebner J E, Siders C W and Barty C P J 2008 Analysis of the scalability of diffraction-limited fiber lasers and amplifiers to high average power *OPT EXPRESS* **16** 13240
[9]     Gray S, Liu A, Walton D T, Wang J, Li M J, Chen X, Ruffin A B, Demeritt J A and Zenteno L A 2007 502 Watt, single transverse mode, narrow linewidth, bidirectionally pumped Yb-doped fiber amplifier *OPT EXPRESS* **15** 17044-50
[10]    Ma P, Zhou P, Ma Y, Su R, Xu X and Liu Z 2013 Single-frequency 332 W, linearly polarized Yb-doped all-fiber amplifier with near diffraction-limited beam quality *APPL OPTICS* **52** 4854
[11]    Eidam T, Wirth C, Jauregui C, Stutzki F, Jansen F, Otto H J, Schmidt O, Schreiber T, Limpert J and Tunnermann A 2011 Experimental observations of the threshold-like onset of mode instabilities in high power fiber amplifiers *OPT EXPRESS* **19** 13218-24
[12]    Smith A V and Smith J J 2011 Mode instability in high power fiber amplifiers *OPT EXPRESS* **19**
[13]    Tao R, Ma P, Wang X, Zhou P and Liu Z 2016 Comparison of the threshold of thermal-induced mode instabilities in polarization-maintaining and non-polarization-maintaining active fibers *J OPTICS-UK* **18** 65501
[14]    Otto H J, Stutzki F, Jansen F, Eidam T, Jauregui C, Limpert J and Tunnermann A 2012 Temporal dynamics of mode instabilities in high-power fiber lasers and amplifiers *OPT EXPRESS* **20** 15710-22
[15]    Shiraki, K., Ohashi and M. 1995 Suppression of stimulated Brillouin scattering in a fibre by changing the core radius *ELECTRON LETT* **31** 668-9
[16]    Liu A 2007 Suppressing stimulated Brillouin scattering in fiber amplifiers using nonuniform fiber and temperature gradient *OPT EXPRESS* **15** 977-84
[17]    Filippov V, Chamorovskii Y, Kerttula J, Golant K, Pessa M and Okhotnikov O G 2008 Double clad tapered fiber for high power applications *OPT EXPRESS* **16** 1929-44
[18]    Alvarez-Chavez J A, Grudinin A B, Nilsson J, Turner P W and Clarkson W A 1999 Mode Selection In High Power Cladding Pumped Fibre Lasers With Tapered Section. In: *Lasers and Electro-Optics, 1999. CLEO '99. Summaries of Papers*

*Presented at the Conference on,* pp 247-8

[19]     Filippov V, Chamorovskii Y, Kerttula J, Kholodkov A and Okhotnikov O G 2009 600 W power scalable single transverse mode tapered double-clad fiber laser *OPT EXPRESS* **17** 1203-14

[20]     Trikshev A I, Kurkov A S, Tsvetkov V B, Filatova S A, Kertulla J, Filippov V, Chamorovskiy Y K and Okhotnikov O G 2013 A 160 W single-frequency laser based on an active tapered double-clad fiber amplifier *LASER PHYS LETT* **10** 65101

[21]     Kerttula J, Filippov V, Chamorovskii Y, Ustimchik V, Golant K and Okhotnikov O G 2012 Principles and performance of tapered fiber lasers: from uniform to flared geometry *Appl Opt* **51** 7025-38

[22]     Kerttula J, Filippov V, Chamorovskii Y, Golant K and Okhotnikov O G 2010 Actively Q-switched 1.6-mJ tapered double-clad ytterbium-doped fiber laser *OPT EXPRESS* **18** 18543-9

[23]     Kerttula J, Filippov V, Chamorovskii Y, Ustimchik V, Golant K and Okhotnikov O G 2013 Tapered fiber amplifier with high gain and output power *LASER PHYS* **22** 1734-8

[24]     Filippov V, Kerttula J, Chamorovskii Y, Golant K and Okhotnikov O G 2010 Highly efficient 750 W tapered double-clad ytterbium fiber laser *OPT EXPRESS* **18** 12499-512

[25]     Zhang H, Du X, Zhou P, Wang X and Xu X 2016 Tapered fiber based high power random laser *OPT EXPRESS* **24** 9112-8

[26]     Filippov V, Ustimchik V, Chamorovskii Y, Golant K, Vorotynskii A and Okhotnikov O G 2015 Impact of Axial Profile of the Gain Medium on the Mode Instability in Lasers: Regular Versus Tapered Fibers. In: *Cleo/europe-Eqec. P. Cj-10.5 1 P. Cj,*

[27]     Zhou Z, Zhang H, Wang X, Pan Z, Su R, Yang B, Zhou P and Xu X 2016 All-fiber-integrated single frequency tapered fiber amplifier with near diffraction limited output *J OPTICS-UK* **18** 65504

[28]     Ran Y, Su R, Ma P, Wang X, Zhou P and Si L 2016 Standard deviation index for stimulated Brillouin scattering suppression with different homogeneities *Appl. Opt.* **55** 3809-13

[29]     Tao R, Ma P, Wang X, Zhou P and Liu Z 2015 1.3   kW monolithic linearly polarized single-mode master oscillator power amplifier and strategies for mitigating mode instabilities *Photonics Research* **3** 86-93

[30]     Johansen M M, Laurila M, Maack M D, Noordegraaf D, Jakobsen C, Alkeskjold T T and Lægsgaard J 2013 Frequency resolved transverse mode instability in rod fiber amplifiers. *OPT EXPRESS* **21** 21847-56

[31]     Ran Y, Tao R, Ma P, Wang X, Su R, Zhou P and Si L 2015 560 W all fiber and polarization-maintaining amplifier with narrow linewidth and near-diffraction-limited beam quality *Appl. Opt.* **54** 7258-63

[32]     Smith A V 2014 Influence of signal bandwidth on mode instability thresholds of fiber amplifiers *Fiber Lasers XII Technology Systems & Applications* **9344**

[33]     Kuznetsov M, Vershinin O, Tyrtyshnyy V and Antipov O 2014 Low-threshold mode instability in Yb3+-doped few-mode fiber amplifiers. *OPT EXPRESS* **22** 29714-25

[34]     Tao R, Ma P, Wang X, Zhou P and Liu Z 2015 Influence of core NA on Thermal-Induced Mode Instabilities in High Power Fiber Amplifiers *LASER PHYS LETT* **12**